\documentclass[12pt]{article}
\usepackage{graphicx}
\usepackage{amssymb}
\usepackage{amsmath}
\usepackage{subfigure}

\def\Acknowledgements{\bigskip  \bigskip \begin{center} \begin{large}
             \bf ACKNOWLEDGEMENTS \end{large}\end{center}}

\begin{document}

\thispagestyle{empty}

\begin{flushright}
Nikhef-2011-009\\
\end{flushright}

\vspace{1.0truecm}
\begin{center}
\boldmath
\large\bf Extraction of the Weak Angle $\gamma$ from $B$ to Charm Decays
\unboldmath
\end{center}

\vspace{0.3truecm}
\begin{center}
Robert Fleischer\,${}^a$ \,and\, Stefania Ricciardi\,${}^b$\\[0.1cm]

\vspace{0.4truecm}

${}^a${\sl Nikhef, Science Park 105, 
NL-1098 XG Amsterdam, The Netherlands}

\vspace{0.2truecm}

\mbox{${}^b${\sl STFC Rutherford Appleton Laboratory,
Chilton, Didcot, Oxfordshire, OX11 0QX, UK}}

\end{center}

\vspace{0.3truecm}

\begin{center}
{\bf Abstract}
\end{center}

{\small
\vspace{0.2cm}\noindent
We give a summary of the discussions in Working Group V of  the CKM2010 workshop
dealing with determinations of the angle $\gamma$ of the unitarity triangle of the 
Cabibbo--Kobayashi--Maskawa matrix from $B$-meson decays into charmed final states. 
}

\vspace{3.2truecm}

\begin{center}
{\sl Summary of Working Group V\\
\mbox{}\\
6th International Workshop\\ on the CKM Unitarity Triangle
(CKM2010)\\
Warwick, United Kingdom, 6--10 September 2010\\
To appear in the Proceedings}
\end{center}

\vfill
\noindent
March 2011

\newpage
\thispagestyle{empty}
\vbox{}
\newpage
 
\setcounter{page}{1}

\section{Introduction}
The angle $\gamma$ of the unitarity triangle (UT) of the Cabibbo--Kobayashi--Maskawa 
(CKM) matrix 
plays a central role for the testing of the flavour sector of the Standard Model (SM). On the one hand,
this angle can be extracted from $B$-meson decays with contributions from loop (penguin)
diagrams. On the other hand, it can also be determined by means of pure tree decays into
final states with charm. The latter avenue was the topic of this working group. We had 12 talks
\cite{zupan}--\cite{poluektov}, with mainly experimental discussions of  time-dependent and 
time-integrated measurements of $\gamma$. 

The outline of this summary report follows closely the working group agenda. In 
Section~\ref{sec:case}, we make the case for a precise measurement of $\gamma$. 
In Sections~\ref{sec:well-est} and \ref{sec:new}, we discuss new results from well-established
methods and from new methods (proposed since the CKM2008 workshop), respectively. 
In Section~\ref{sec:prop}, we address the prospects for the measurement of $\gamma$
at the LHCb experiment and beyond. Finally, we summarize our concluding remarks in
Section~\ref{sec:concl}. For references to original papers, the reader is referred to the
contributions listed in \cite{zupan}--\cite{poluektov}.

\boldmath
\section{The Case for a Precise Measurement of $\gamma$}\label{sec:case}
\unboldmath
Decays of $B$ mesons into final states with charm offer a variety of strategies to 
determine $\gamma$. Here the sensitivity on this angle arises from interference
effects between $\bar b\to \bar c$ and $\bar b\to\bar u$ quark-level processes in decays of the kind
$B\to DK$. These decays originate only from tree-diagram-like topologies, i.e.\
we have no contributions from penguin diagrams, and involve only one weak phase difference. 
In addition to $\gamma$, also other $B$ and $D$ hadronic parameters enter the
analyses, typically involving the ratio of colour-suppressed to colour-allowed
amplitudes and CP-conserving strong phases. Several methods to extract 
all unknown parameters from the data were proposed, using combinations of several 
$D$ modes or input from charm studies. 

In his talk \cite{zupan}, setting the stage for the discussions in our Working Group,
Jure Zupan addressed the question: how clean are these determinations? The
usual kind of reasoning is along the lines that these decays involve only tree-level
amplitudes and are hence not affected by theoretical uncertainties. Moreover, they
are considered to received no new-physics (NP) contributions and serve as standard 
candles for the SM. 

A potential source for theory errors is $D^0$--$\bar D^0$ mixing. 
Since this is a 
strongly suppressed phenomenon in the SM, it leads 
to a small error for the determination of $\gamma$, at most $\sim1^\circ$ if it is completely
neglected. However, if the $D^0$--$\bar D^0$ mixing parameters $x_D$ 
($\propto$ mass difference) and $y_D$ ($\propto$ lifetime difference) are measured 
precisely, the mixing effects can be taken into account. Consequently,  $D^0$--$\bar D^0$ 
mixing does not lead to an irreducible uncertainty. 

The second potential source of theoretical 
uncertainty is related to electroweak corrections, which may change the phase structure of 
the decay amplitudes through box-diagram topologies. However, the corresponding irreducible 
theoretical error is estimated as $\delta\gamma/\gamma={\cal O}(10^{-6})$, and is hence not any
issue from the practical point of view \cite{zupan}. 

Concerning the current status of the measurement of $\gamma$ through $B$-meson
decays into charmed final states, the error is still very large:
\begin{equation}
\gamma=\left\{
\begin{array}{ll}
\left(71^{+21}_{-25}\right)^\circ & \mbox{(CKMfitter Collaboration)}\\
(73\pm11)^\circ & \mbox{(UTfit Collaboration).}
\end{array}
\right.
\end{equation}
Interestingly, individual measurements are more precise than the CKMfitter average, which
raises questions on the statistical procedures used. This was not discussed in detail within
our working group. However, there was consensus that a larger data sample will reduce the
disagreement between different statistical treatments. 

The current experimental precision is much beyond the $B$-factories design expectations, which
is mainly due to the excellent performance of the accelerators and experiments and to the use of the
GGSZ method, which entered the scene at the CKM2003 workshop. 

Nevertheless, the determinations of $\gamma$ from $B$ decays into charmed final states
have the smallest theoretical uncertainties, but suffer from the largest experimental errors
among all constraints for the UT. Let us next discuss new results from well-established
methods in more detail.

\section{New Results from Well-Established Methods}\label{sec:well-est}
Well-established methods exist for two distinct types of analyses: time-integrated, which access $\gamma$ through measurements of direct CP violation in charged or self-tagged neutral $B$ decays, and time-dependent, which extract the weak-phase $2\beta_{(s)}+\gamma$ from measurements of interference between $B^0_{(s)}$ decays with and without mixing. 
\subsection{Time-Integrated Measurements}
Time-integrated measurements have been successfully performed by the $B$-factory experiments and currently provide the best precision on $\gamma$. They are limited by statistical uncertainties, hence the present knowledge of $\gamma$ can be significantly improved by additional measurements with larger $B$ samples. It should be noted that the $B$-decay parameters do not depend on the $D$-decay mode (and vice-versa), therefore experiments can gain more than just statistics by extracting $\gamma$ via a combined fit to different channels.
Different $D$ decay channels have been exploited by the Belle and BaBar collaborations, 
corresponding to three different time-integrated methods: the GLW method for CP eigenstates, the ADS method for doubly Cabibbo-supppressed decays, and the GGSZ  method for three-body self-conjugate final stats.

Particularly interesting new results have been presented at this workshop for the ADS method for $B\to D(K\pi) K$ decays for both the Belle~\cite{horii} and the CDF~\cite{squillacioti} collaborations. This is a powerful method where the CP asymmetry is enhanced for final states with two opposite-charge kaons, because the two interfering amplitudes have similar size. 

At CKM2010, Belle has reported the first evidence of the suppressed decay 
$B^-\to D(K^+\pi^-)K^-$,\footnote{Charge conjugation is implied everywhere, unless otherwise stated.} obtained from the full data sample collected at the $\Upsilon(4S)$ resonance, corresponding to $7.7\times 10^8$ $B\bar{B}$ pairs. 
The preliminary results on the relative rate, $\cal{R}_{\rm ADS}$, of $B^-\to D_{\rm DCS}K^-$ 
to $B^-\to D_{\rm CF}K^- $, 
\begin{equation}
{\cal{R}}_{\rm ADS} = \left[1.62 \pm 0.42\mbox{(stat.)}^{+0.16}_{-0.19}\mbox{(syst.)}\right] 
\times 10^{-2},
\end{equation}
corresponds to $3.8~\sigma$ evidence for the suppresssed mode, while no significant CP asymmetry, 
\begin{equation}
{\cal{A}}_{\rm ADS} = -0.39 \pm 0.26\mbox{(stat.)}^{+0.06}_{-0.04}\mbox{(syst.)},
\end{equation}
is observed between the $B^+$ and $B^-$ suppressed decays.

Time-integrated methods are well-suited for hadron colliders, because they do not require $B$-tagging, hence their large $B$ production can be fully exploited. However, selecting pure samples of fully hadronic $B\to DK$ decays requires detectors with excellent trigger and PID capabilities. At this workshop, the CDF Collaboration presented the first measurement of ${\cal{R}}_{\rm ADS}$ and ${\cal{A}}_{\rm ADS}$ at the Tevatron. Results are based on a luminosity of 5~fb$^{-1}$ and are in good agreement with existing $B$-factory measurements. These results supplement the recently published GLW analysis by CDF within a global programme to measure $\gamma$ from tree-dominated processes. 
The precision of these measurements is comparable, even if not competitive yet, with the current best measurements from the $B$ factories, and will improve with the full data sample of 
(10--12)~fb$^{-1}$, which is expected by the end of 2011. Most importantly, these results are a demonstration of the feasibility of these measurements at hadron colliders.

The BaBar Collaboration has recently published new results for all the three methods (ADS, GLW and GGSZ)~\cite{marchiori}.
The measurements have been performed on the full sample of 468 million $B\bar{B}$ pairs. The achieved precision on $\gamma$, around 15$^{\circ}$, is dominated by the result achieved with the GGSZ method. The BaBar GGSZ analysis is based on $B^\pm\to DK^\pm$, $D^*K^{\pm}$, and $DK^{*\pm}$ decays, followed by neutral $D$-meson decays to $K^0_{\rm S}h^+h^- (h=\pi,K)$. The weak phase $\gamma$ and other $B$-decay parameters are extracted from an amplitude fit to the Dalitz plot distributions of the $D$ decays. The $D^0$ and $\bar{D^0}$ decay amplitudes to $K^0_{\rm S} h^+h^-$ are modeled by the coherent sum of a non-resonant part and several intermidiate two-body decays that proceed through $K^0_{\rm S}h$ or $h^+h^-$ resonances.
The uncertainties in the model introduce an additional systematic error on $\gamma$. However, the result 
\begin{equation}
\gamma (\mathrm{mod} 180^{\circ}) = \left[68 \pm 14\mbox{(stat.)} \pm 4\mbox{(syst.)} \pm 3\mbox{(model)}\right]^{\circ}
\end{equation}
is still dominated by the statistical error. It can be compared to the Belle GGSZ result~\cite{horii} 
\begin{equation}
\gamma (\mathrm{mod} 180^{\circ}) = \left[78.4^{+10.8}_{-11.6}\mbox{(stat.)} \pm 3.6\mbox{(syst.)} \pm 8.9\mbox{(model)}\right]^{\circ},
\end{equation} 
which is obtained using $D\to K^0_{\rm S}\pi\pi$ decays only, but a larger data sample and a less sophisticated decay model. 

The model error is hard to quantify, but can be avoided by using a model-independent approach. In the latter, the model dependence is lifted by relating the $B$ yields to discrete measurements of the strong-phase difference, $\Delta\delta_D$, between $D^0$ and $\bar{D^0}$ to $K^0_{\rm S}hh$, in bins of the Dalitz plot. Only experimental observables are involved in this case, hence there are no model uncertainties.  
These measurements have been recently performed for different binning choices and for both $D\to K^0\pi\pi$ and $D\to K^0KK$ at CLEO-c~\cite{ricciardi} with the full data-sample of quantum-correlated $D\bar{D}$ decays collected at the $\psi(3770)$, which corresponds to 818 pb$^{-1}$. The model-independent approach is expected to suffer from a small loss in statistical precision compared to the model-dependent one, which is unbinned, hence makes optimal use of all available information. This loss has been estimated to be about 10\%. The CLEO-c uncertainties on the strong-phase parameters will also induce a systematic uncertainty on $\gamma$, when a model-independent approach is adopted. This has been evaluated to be about (3--4)$^\circ$  for $B^\pm\to~D(K^0_{\rm S} K^+K^-)K^\pm$, and (2--4)$^\circ$ for $B^\pm\to~D(K^0_{\rm S}\pi^+\pi^-)K^\pm$, which varies within the given range according to the binning choice. These small residual errors, which are due mainly to the limited size of the CLEO-c data sample, are adequate for the precision on $\gamma$ which is expected at LHCb with the GGSZ method, and could be further reduced with larger data samples. 

Other CLEO-c analyses have been presented at this workshop~\cite{ricciardi}, which allow for a significant improvement in the precision for $\gamma$ from time-integrated measurements of $B\to DK$ decays. These include a new preliminary result on the strong-phase difference $\Delta\delta_D$ in $D\to K\pi$, and the results on the coherence factor and on $\Delta\delta_D$ in the multibody decays, $D\to K\pi\pi^0$ and $D\to K\pi\pi\pi$.  
Prospects for improving these and other measurements of the $D$-decay parameters at future charm--tau factories, and their impact on the measurement of $\gamma$, have been discussed by Spradlin~\cite{spradlin} at this workshop. One clear message emerges: the contribution of physics at the charm threshold is invaluable to the precise measurement of $\gamma$.

\subsection{Time-Dependent Measurements}

Another well-established method to determine $\gamma$ is by exploiting the interference between 
$\bar{b}\to \bar{c}$ and $b\to u$ mediated transitions that is caused by $B^0$--$\bar{B^0}$ 
mixing and occurs when both the $B^0$ and the $\bar{B^0}$ mesons can decay to a 
common final state. A time-dependent analysis is required and is sensitive to the sum of the mixing phase, $2\beta$, and the relative phase between the $B^0$ and $\bar{B^0}$ decay amplitudes, 
$\gamma$. 

Abundant decays, such as $B^0\to D^{(*)\mp}\pi^\pm$, can be used.
In this case, the $b\to u$ decay amplitude is doubly Cabibbo-suppressed, while the $\bar{b}\to \bar{c}$ transition is Cabibbo-favoured. Therefore, the magnitude of the ratio between the two, 
\begin{equation}
r = \left|\frac{A(\bar{B^0}\to D^{(*)-}\pi^+)}{A(B^0\to D^{(*)-}\pi^+)}\right|, 
\end{equation}
which governs the size of the CP violation effect, is expected naively to be about 2\%. Because of this very small value, external input on $r$ is required to extract the weak phase. Discussions at this workshop have focussed on these external measurements and the associated systematic uncertainties. 

The BaBar Collaboration has estimated $r$ from the ratio of the branching fractions of  
$B^0\to D^{*+}_s \pi^-$  and $B^0\to D^{*-}\pi^+$ as follows \cite{rubin}:
\begin{equation}
r = \sqrt{\frac{\mathcal{B}( B^0\to D^{*+}_s \pi^-)}{\mathcal{B}(B^0\to D^{*-}\pi^+)}}
\frac{f_{D^*}}{f_{D^*_s}} \tan(\theta_{\rm C}) = 0.015^{+0.004}_{-0.006} \times (1.0 \pm 0.3),
\end{equation} 
where the 30\% systematic uncertainty accounts for possible non-factorizable,
$SU(3)$-breaking corrections. Time-dependent measurements have been performed using 232 million $B\bar{B}$ pairs, which is about half of the full data-sample.
By combining results from fully-recontructed $B^0\to D^{(*)-} \pi^+$, fully-reconstructed 
$B^0\to D^{-(*)} \rho^+$  
and partially-reconstructed $B^0\to D^{*-} \pi^+$ decays, BaBar obtains $|\sin(2\beta+\gamma)|>0.64$ (0.40) at the 64\% (90\%) confidence level. 

The Belle Collaboration has performed the analysis using fully-reconstructed $B^0\to D^{(*)-}\pi^+$ events from 386 million $B\bar{B}$ pairs, and partially recontructed $B^0\to D^{*-}\pi^+$ events from a larger data-sample of 657 million $B\bar{B}$ pairs. Results can be found in Ref.~\cite{onuki}.
A new measurement for the $D\pi$ final state has been presented for the first time at this conference. 
Using $SU(3)$ flavour-symmetry assumptions, Belle obtains 
\begin{eqnarray}
r_{D\pi} & = &[1.71 \pm 0.11 (\mathrm{stat.}) \pm 0.09 (\mathrm{syst.}) \pm 0.02 (\mathrm{th.})]\% \\
r_{D^*\pi} &=& [1.58 \pm 0.15 (\mathrm{stat.}) \pm 0.10 (\mathrm{syst.}) \pm 0.03 (\mathrm{th.})]\%.
\end{eqnarray} 
These are the most precise determination of $r$, but possible non-factorizable $SU(3)$-breaking effects are not completely accounted for in the theory error.
Another promising way to determine $r$ is through the following isospin relation: 
\begin{equation}
r_{D^*\pi} = \sqrt{\frac{2{\mathcal B}( B^+\to D^{*+} \pi^0)}{{\mathcal B}(B^0\to D^{*-}\pi^+)}\frac{\tau_{B^0}} {\tau_{B^+}}}. 
\end{equation}
In this case the measurement is limited by statistics, as only an upper limit is available for the branching fraction of the $B^+\to D^{*+} \pi^0$ decay. With his method Belle finds  
$r_{D^*\pi}<0.051$ (90\% C.L.).

BaBar has exploited also $B$ decays that can exhibit larger interference effects, hence a larger value of $r$.  These include 
$B^0 \to D^\mp K^0\pi^\pm$, where a time-dependent Dalitz-plot analysis has been performed with 347 million $B\bar{B}$ pairs.
The values of $2\beta+\gamma$ is obtained as a function of $r$.
For $r= 0.3$, the result 
\begin{equation}
2\beta+\gamma (\mathrm{mod} 180^\circ) =  (83 \pm 53 \pm 20)^{\circ}
\end{equation}
is obtained, where the central value has a weak dependence on $r$. 

\section{New Methods} 
\label{sec:new}
Since 2008, a few new methods to measure $\gamma$ have been proposed. Among these, the multibody 
$B^0\to D K^+ \pi^-$ analysis is particularly well-suited for LHCb~\cite{zupan}, 
because $\gamma$ can be extracted by reconstructing $D$ decays with only charged particles in the final state. 
In addition, the Dalitz plot of this $B$ decay features a flavour-specific resonant decay $D^{*-}_2(2460)\to \bar{D^0} \pi^-$. 
The interference of $D^{*-}_2(2460) K^+$ with other resonances in 
the $B^0\to D K^+ \pi^-$ Dalitz plot, such as  
$D K^{*0}(892)$, allows $\gamma$ to be extracted with better sensitivity compared to that estimated for 
the quasi two-body $B^0\to D K^{*0}$ decay with a GLW/ADS method.
Similarly to the quasi two-body determination, the multibody method requires
the reconstruction of $D$ decays to CP eigenstates and flavour-specific modes, but 
is based on the determination of relative decay amplitudes, rather than the determination of decay rates.
The expected precision achievable at LHCb with this method has been recently re-evaluated using data yields 
extrapolated from the 2010 data-taking. 
Approximately 700 $B^0\to D K^+\pi^-$ decays followed by the favoured $D\to K^+\pi^-$
are expected in 1~fb$^{-1}$, from which $\gamma$ can be determined 
with a statistical uncertainty of about 20$^\circ$~\cite{williams}.

Another interesting channel is $B^0_s\to J/\psi K_{\rm S}$, which has been observed by CDF 
in the summer of 2010 and will be of interest for the LHCb experiment \cite{fleischer}.
This channel is caused by $\bar b\to \bar cc \bar d$ quark-level transitions and is the 
$U$-spin partner of the ``golden" decay $B^0_d\to J/\psi K_{\rm S}$. Thanks to the interference
between tree and penguin topologies, which are not doubly Cabibbo-suppressed as in
$B^0_d\to J/\psi K_{\rm S}$, the UT angle $\gamma$ can be determined through measurements
of the CP-violating asymmetries of $B^0_s\to J/\psi K_{\rm S}$ and the application of the $U$-spin
symmetry. A first feasibility study for LHCb shows that experimental sensitivity at the 
few-degree level can be obtained with an upgraded LHCb detector (see next section). Although 
interesting on its own, this determination of $\gamma$ is not competitive in terms of precision
with pure tree strategies. However, the measurement of CP violation in $B^0_s\to J/\psi K_{\rm S}$ 
allows also to determine hadronic penguin parameters and to control their impact on the extraction of the $B^0_d$--$\bar B^0_d$ mixing phase $\phi_d$ through a measurement of the mixing-induced CP violation in $B^0_d\to J/\psi K_{\rm S}$. As discussed in \cite{fleischer}, 
this will be the major application of $B^0_s\to J/\psi K_{\rm S}$ at LHCb. 
Such an analysis is actually needed in order to fully exploit LHCb's impressive experimental precision for the determination of $\phi_d$ from $B^0_d \to J/\psi K_{\rm S}$, and will be 
an interesting study at an LHCb upgrade, which may eventually allow us to resolve NP effects
in $B^0_d$--$\bar B^0_d$ mixing.

\section{Prospects: LHCb and Beyond}\label{sec:prop}

The $B^0_s$ system provides additional opportunities for the precise measurement of $\gamma$. Among these, the time-dependent measurement with 
$B_s\to D_s^\pm K^\mp$ decays is particularly promising. 
This measurement is unique to LHCb, because only LHCb has both access to a large $B^0_s$ data-sample and the ability to resolve the fast $B^0_s$--$\bar B^0_s$ oscillations.
The tree-level sensitivity to $2\beta_s +\gamma$ arises from the interference between the decay with and without mixing.
The value of $2\beta_s +\gamma$ can be converted to a measurement of $\gamma$ because $\beta_s$ will be well-determined by measurements of $B^0_s\to J/\psi\phi$ decays. 

This analysis  presents some advantages in comparison with the time-dependent measurements from $B^0_d$ decays. 
It benefits from the expected sizeable width difference, $\Delta\Gamma_s$, in the $B^0_s$ system, which provides additional sensitivity to $\gamma$ through the inclusion of untagged events. It also benefits 
from the large interference in these decays, given that the ratio of the magnitude between the two interfering amplitudes
is expected to be approximately 0.4, which is large enough for it to be determined from data.
By performing a simultaneous fit to $B^0_s\to D_s K$ and $B^0_s\to D_s \pi$ events, all physical unknowns and experimental parameters can be extracted from data, so $\gamma$ can be measured without theoretical uncertainties. 
In 2011, with 1~fb$^{-1}$, it is expected that the first measurements of the CP-violating observables will be performed~\cite{gligorov}. However, an integrated luminosity of about 2~fb$^{-1}$ at 7~TeV is required for an unambiguous extraction of $\gamma$ from this mode. The achievable precision depends on the flavour tagging performance,
but should be competitive with the LHCb results from the time-integrated measurements.

In addition to colour-allowed $B^0_s\to D_s^\pm K^\mp$ decays, also colour-suppressed $B^0_s\to D\phi$ decays
offer sensitivity to $\gamma$ at tree-level. In this case, a time-integrated measurement of $\gamma$ exploiting untagged decays has been proposed~\cite{zupan}. The untagged method allows LHCb to exploit fully its statistical power and to mitigate the effect of the small branching fraction for this mode. If a sufficient number of different $D$ decays are used, there is enough experimental information to extract $\gamma$ and all the hadronic parameters. Sensitivity studies performed at LHCb have evaluated that a statistical uncertainty of about 20$^{\circ}$ on $\gamma$ can be achieved by this method with 1~fb$^{-1}$. This level of precision is comparable to the LHCb expectations for the other time-integrated methods.

The long-term prospects for the determinations of $\gamma$ from $B_{(s)}\to D_{(s)} K$ decays 
were discussed by Anton Poluektov \cite{poluektov}. Recently, two next generation $e^+e^-$
facilities have been approved: the SuperB project in Italy and SuperKEKB in Japan. SuperB 
has a design luminosity of $10^{36}\,\mbox{cm}^{-2}\,\mbox{s}^{-1}$  and aims at 
$75\,\mbox{ab}^{-1}$, which is very similar to the goals of SuperKEKB, with 
$8\times 10^{35}\,\mbox{cm}^{-2}\,\mbox{s}^{-1}$  and a goal of an integrated luminosity of
$50\,\mbox{ab}^{-1}$. Both projects hope to start operation around 2015 and should take data
until 2020. Concerning LHCb, a data sample of (6--7)\,$\mbox{fb}^{-1}$ is expected to be 
collected by 2016. There are plans to upgrade LHCb afterwards to run with an increased
luminosity of up to $2\times 10^{33}\,\mbox{cm}^{-2}\,\mbox{s}^{-1}$, which would result in an 
expected data sample of (50--100)\,$\mbox{fb}^{-1}$ by 2020.

In \cite{poluektov}, a detailed comparison of the $\gamma$ reach for SuperB 
(50\,$\mbox{ab}^{-1}$) and an upgraded LHCb experiment (50\,$\mbox{fb}^{-1}$) was given. 
The bottom line is that various independent methods should allow measurements of $\gamma$ 
with uncertainties at the $(1\mbox{--}3)^\circ$ level, using ADS and GLW methods, 
$B\to D(K_{\rm S}\pi\pi)K$ Dalitz analyses, self-tagging $B^0\to DK\pi$ modes, and the 
$B_s\to D_s K$ channel, which is only accessible at LHCb. The overall sensitivity at the
LHCb upgrade looks potentially better than that of SuperB, although potential backgrounds
can reduce it. On the other hand, SuperB looks more stable with respect to ``unfortunate" 
parameter combinations, where the sensitivity can be significantly reduced. 

In order to fully exploit the $B$-decay data, it is desirable to have a large data sample, i.e.\
$\sim(10\mbox{--}20)\,\mbox{fb}^{-1}$, at the charm threshold. Such data could be recorded at BES-III,
a new dedicated charm-tau factory, or running SuperB at low energy. 

\section{Concluding Remarks}\label{sec:concl}
The determination of $\gamma$ from $B$ decays into charmed final states has continued
to progress. By the next CKM workshop, we expect new exciting experimental
results from $B$-decay studies at hadron colliders. At this workshop, CDF has demonstrated
that analyses of $B\to D K$ modes are possible at hadron colliders. The corresponding
new ADS/GLW results with 5\,$\mbox{fb}^{-1}$ are roughly competitive with those of the $B$ factories, and
the exploration continues, with an expected data set of $(10\mbox{--}12)\,\mbox{fb}^{-1}$ by
the end of the Tevatron run in 2011. 

Concerning LHCb, we look forward to a variety of interesting measurements, with an 
expected uncertainty $\Delta\gamma<10^\circ$ by the end of 2012. The excellent tracking,
PID and trigger performance for these multi-hadron decay modes demonstrated with the
$<1\,\mbox{pb}^{-1}$ of data analyzed in the summer of 2010 give us confidence that this
can actually be achieved. The precise determination of $\gamma$ continues to be a key 
target of the $B$-physics programme!

\Acknowledgements
We would like to thank all participants of our working group for excellent talks and discussions,
as well as the organizers of CKM2010 for doing such a great job in hosting this workshop at
the University of Warwick.


\begin{thebibliography}{99}

\bibitem{zupan}J.~Zupan,
  ``The case for measuring $\gamma$ precisely,''
  arXiv:1101.0134 [hep-ph].

\bibitem{rubin}A. Rubin  [BaBar Collaboration],   ``Time-dependent $\gamma/\phi_3$ 
measurements by Babar,'' arXiv:1102.3427 [hep-ex].

\bibitem{onuki}Y.~Onuki [Belle Collaboration],
  ``Belle time-dependent $\gamma$ measurements,''
  arXiv:1012.2942 [hep-ex].

\bibitem{gligorov} V.~V.~Gligorov [LHCb Collaboration],
  ``Time dependent measurements of the CKM angle $\gamma$ at LHCb,''
  arXiv:1101.1201 [hep-ex].
  
\bibitem{ricciardi}S.~Ricciardi [CLEO Collaboration],
  ``CLEO-c inputs to the determination of the CKM angle $\gamma$,''
  arXiv:1101.4855 [hep-ex].

\bibitem{horii} Y.~Horii [Belle Collaboration], 
  ``Belle time-integrated $\gamma/\phi_3$ measurements,''
  arXiv:1101.0878 [hep-ex].
  
\bibitem{marchiori}G.~Marchiori [BaBar Collaboration],  
``Time-integrated measurements of the CKM angle $\gamma/\phi_3$ in BaBar,''
  arXiv:1012.4979 [hep-ex].
  
\bibitem{squillacioti}P.~Squillacioti  [CDF Collaboration],
  ``Time-integrated measurements of $\gamma$ at the Tevatron and prospects,''
  arXiv:1012.1781 [hep-ex].
  
\bibitem{williams}M.~Williams  [LHCb Collaboration],
  ``Time-integrated measurements and prospects for the CKM angle $\gamma$ at
  LHCb,''
  arXiv:1101.1190 [hep-ex].

\bibitem{spradlin}P.~Spradlin, ``Charm inputs for the next decade".

\bibitem{fleischer}K.~De Bruyn, R.~Fleischer and P.~Koppenburg,
  ``Extracting gamma and Penguin Parameters from $B^0_s \to J/\psi K_{\rm S}$,''
  arXiv:1012.0840 [hep-ph].
  
\bibitem{poluektov}A.~Poluektov,
  ``Ultimate sensitivity on $\gamma/\phi_3$ from $B\to DK$,''
  arXiv:1101.4592 [hep-ex].


\end{thebibliography}
\end{document}